\documentclass[superscriptaddress]{revtex4}

\usepackage{epsf}
\newcommand{\be}{\begin{equation}}
\newcommand{\ee}{\end{equation}}
\newcommand{\bea}{\begin{eqnarray}}
\newcommand{\eea}{\end{eqnarray}}
\renewcommand{\t}{\tau}
\newcommand{\sgn}{\mbox{sign}}
\newcommand{\tri}{\mbox{tri}}

\begin{document}

\title{Efficiency through disinformation}
\author{Richard Metzler}
\affiliation{New England Complex Systems Institute,
 24 Mt. Auburn St., Cambridge, MA 02138, USA}
\affiliation{Department of Physics, Massachusetts Institute of Technology,
 Cambridge, MA 02139, USA}
\author{Mark Klein}
\affiliation{Center for Coordination Science, Massachusetts Institute of Technology,  Cambridge, MA 02139, USA}
\author{Yaneer Bar-Yam}
\affiliation{New England Complex Systems Institute,
 24 Mt. Auburn St., Cambridge, MA 02138, USA}

\begin{abstract}
We study the impact of disinformation on a model of resource allocation
with independent selfish agents: clients send
requests to one of two servers, depending on which one is perceived as
offering shorter waiting times. Delays in the information about the
servers' state leads to oscillations in load. Servers can give 
false information about their state (global
disinformation) or refuse service to individual clients 
(local disinformation). We discuss the tradeoff between 
positive effects of disinformation (attenuation of
oscillations) and negative effects (increased fluctuations and reduced
adaptability) for different parameter values.
\end{abstract}

\maketitle
Competition for limited resources occurs in many different situations, and
it often involves choosing the resource least popular among competitors --
one can think of drivers who want to take the least crowded road, investors
who want to buy hot stocks before other buyers drive the price up, 
computers that send requests to the least busy servers, and many more. 
From an individual perspective, agents in these scenarios act selfishly --
they want to achieve their particular aims. At the same
time, this selfish behavior can be beneficial for the system as a whole,
insofar as it leads to effective resource utilization 
\cite{Smith:Inquiry}; however, this is not always the case. 
From the point of view of the
system, the problem then becomes one of distribution of resources, rather than
competition.

The most commonly
studied model in this context, the Minority Game (MG)
\cite{Challet:Emerg.,Challet:Phase,MGHomepage}, has agents choose one of
two alternatives, basing their decision on a short history of the global
state of the system. A multitude of possible strategies for the agents can be
conceived. One recurring theme in many of the MG's
variations is oscillations of preferences: in some cases, preferences
oscillate in time \cite{Nakar:Semianalytical,Reents:Stochastic}, whereas in
others, a reaction of the system to a given history pattern will be followed
by the opposite of that reaction the next time this pattern occurs 
\cite{Marsili:Trading,Savit:Comp.}. The presence of oscillations 
indicates suboptimal resource utilization. 
Their source is the fact that agents make
their decision based on obsolete data, i.e., there is a delay 
between the time that the
information underlying their decision is generated and the time their
decision is evaluated. This is often obscured by the use of discrete time
steps in most variations of the MG.
In this paper, we study a continuous-time MG-like scenario with an explicit
time delay,  which was inspired by the competition of computers 
for network server service, but can serve as a model case for other problems.

We also introduce a new way to think about controlling the dynamics 
of the system. In previous papers, possible ways to improve efficiency  
were explored from the point of view of agents' strategies: how should
an agent behave to achieve maximum payoff? The result, however, was
measured as an aggregate quantity -- the total degree of resource utilization.
In this paper, we  assume that the agents are selfish and short-sighted, 
and their strategy is not accessible to modification. 
Responsibility for system efficiency lies with the
servers, who can influence behavior by providing incorrect information. 
We first introduce the system, study its native 
dynamics, and determine under
what circumstances control measures can improve efficiency. 
We then present various possible scenarios of influencing
the global behavior.

{\em The model --} The system we consider \cite{Klein:Handling} consists of 
two servers $R_1$ and $R_2$, which offer the same service to a number $N$ of
clients. Clients send data packets to one of the servers. After a travel time
$\t_T$, the packets arrive at the server, and are added to a queue. 
Servers need a time $\t_P$ to process each request. We choose the time
scale such that $t_P=1$. When a client's request
is completed, the server sends a ``done'' message (which takes another
$\t_T$ to arrive) to the client. The client is then idle for a time $\t_I$,
after which it sends a new packet.  
Clients receive information about the state of each 
server. They decide which server offers shorter waiting times based 
on this information,
and send their packets to the respective server. However, for various
reasons, the information they receive is obsolete -- they have access to the
length of the queues a delay time $\t_D$ ago.

The system can be solved simply, if both servers accept all incoming requests 
and demand is distributed uniformly enough, 
such that both servers are busy at all 
times. The only relevant variables are $N_1(t)$ and $N_2(t)$, the number
of clients whose data is in the queue or being processed by $R_1$ and
$R_2$, respectively, at time $t$; we treat them as continuous variables.
Idle clients do not have to be taken into account explicitly; neither do
clients who are waiting for their ``done'' message from the server -- for
our purposes, they are the same as idle agents. We will first solve the 
problem neglecting agents whose message is travelling to the server, then
include non-vanishing travel times.

There are only two processes which change the length of the queues: 
(a) Due to processed requests, both $N_1$ and $N_2$ decrease by 1 per time
unit. 
(b) In the same time span, 
two clients (whose data was processed by $R_1$ and $R_2$
a time $\t_T+\t_I$ ago) compare the obsolete values $N_1(t-\t_D)$ and
$N_2(t-\t_D)$ and add their requests to the queue according to this
information.
We write delay-differential equations for $N_i$:
\bea
\frac{dN_1}{dt} &=& 2\Theta(N_2(t-\t_D)- N_1(t-\t_D)) -1; \nonumber \\
\frac{dN_2}{dt} &=& 2\Theta(N_1(t-\t_D)- N_2(t-\t_D)) -1,
\eea
where $\Theta$ stands for the Heaviside step function.
This can be simplified even more by introducing $A(t) = N_1(t)-N_2(t)$, the
difference in queue lengths:
\be
\frac{dA}{dt} = -2\; \sgn(A(t-\t_D)). \label{DEL-diffeq1}
\ee
This has a steady-state solution
\be
A(t) = 2\, \t_D\,\tri\left(\frac{t}{4\t_D} + \phi\right), \label{DEL-steady1}
\ee
where $\tri(x)$ is the triangle function
\be
\tri(x) = \left\{\begin{array}{ll}
      4x-1&\mbox{~~~~for~~~~}0\leq x <1/2, \\
      -4(x-1/2)+1 &\mbox{~~~~for~~~~} 1/2 \leq x <1, \mbox{~~~periodic in 1,}
\end{array}  \right.
\ee
and $\phi$ is a phase determined by initial conditions.

\begin{figure}
  \epsfxsize= 0.9 \columnwidth
  \epsffile{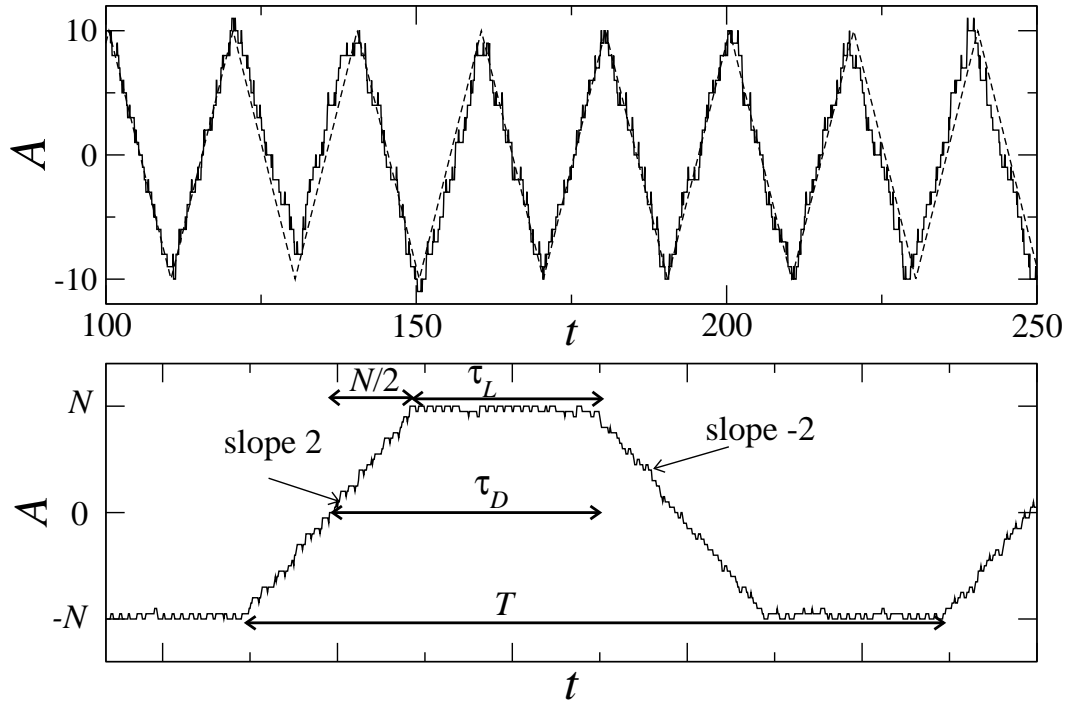}
  \caption{Unmodified dynamics of the system. {\bf Top}: Comparison between
    simulations with $N=100$ and $\t_T=0$ to Eq. (\ref{DEL-steady1}). The
    delay is $\t_D=5$, yielding a period of $20$ and an amplitude of $10$.
    {\bf Bottom}: An example of $A(t)$ in the regime where servers go idle
    periodically.}
  \label{DEL-norejectfig}
\end{figure}

Eq. (\ref{DEL-steady1}) shows that the solution is oscillatory.
The frequency of oscillation 
is only determined by the delay, and the amplitude by the ratio 
of delay time to processing time -- the total number of clients does not
play a role. Clients typically 
spend much of their time with their request in the
queue, and adding more clients only makes both queues longer. Also, 
if the delay goes to zero, so does the amplitude
of oscillations: the minority game is trivial if agents can instantaneously
and individually respond to the current state.
Fig. \ref{DEL-norejectfig} shows that computer simulations are in good
agreement with Eq. (\ref{DEL-steady1}) and in particular 
that the treatment of $N_i$ as
continuous variables works well even for small amplitudes. 

Introducing a non-vanishing travel time $\t_T$ has the same effect on $A(t)$ as
increasing the delay time: it leads to
the delay-differential equation 
\be
\frac{dA}{dt}(t+\t_T) = -2 \sgn(A(t-\t_D)). 
\ee
The solution is given by Eq. (\ref{DEL-steady1}) with $t_D$ replaced by 
$t_D+t_T$.

{\em The impact of idle servers --} The case where oscillations become so 
strong that servers go idle periodically ($2\t_D>N$) can be treated in a
similar framework, for $t_T=0$: 
once the difference in queue lengths reaches the value $\pm N$,
one queue
ceases to process requests. Hence, the rate of requests at the other
server drops from $2$ to $1$ -- exactly the rate at which it keeps
processing them. The queue length at the active server therefore stays
constant for some time $\t_L$.
An example of the resulting curve can be seen in Fig. \ref{DEL-norejectfig}
(bottom). 
Starting from the time where $A(t)$ crosses the zero line, it will take a
time $\t_D$ for clients to realize that they are using the ``wrong''
server, so $\t_D = \t_L + N/2$, or $\t_L=\t_D-N/2$. 
The period $T$ of the oscillations
is then $T=2 \t_L+ 2 N = 2\t_D + N$, which is smaller than $4\t_D$. 
Data throughput of the system drops
from $2$ to $1 + N/(2 \t_D)$ (in units of $1/\t_P$). All of
this is again in good agreement with simulations.


The results above specify the system parameters for which oscillations
affect throughput, and how strong the impact is. 
We now consider ways for the
servers to reduce oscillations. Two methods suggest themselves: 
global disinformation from both servers and individual rejection by 
each server.

{\em Global disinformation --} If the servers have control 
over the information that clients receive on
the servers' status, they can intentionally make the information
unreliable. Let us assume clients have a
probability $p$ of receiving the wrong answer, and accordingly choose the
``wrong'' server. The update equations are:
\bea
\frac{dN_1}{dt} &=& 2[(1-p) \Theta(-A(t-\t_D)) +
p\; \Theta(A(t-\t_D))] -1; \nonumber \\
\frac{dN_2}{dt} &=& 2[(1-p)\Theta(A(t-\t_D)) +
p\; \Theta(-A(t-\t_D)) ] -1,
\eea
leading to 
\be
\frac{dA}{dt} = -2 (1-2 p) \sgn(A(t-\t_D)).\label{DEL-diffeq2}
\ee
This equation has the form of Eq. (\ref{DEL-diffeq1}) with a prefactor
of $1-2p$, and has a steady-state solution 
\be
A(t) = 2\, \t_D\,(1-2p)
 \tri\left(\frac{t}{4\t_D} + \phi\right), \label{DEL-steady2}
\ee
for $p<1/2$. At $p=1/2$, no information is available: clients' decisions
are random, and queue lengths perform a random walk, whose fluctuations
are not captured by the deterministic framework we are using. Even for values
$p<1/2$, fluctuations may become larger than the typical amplitude of
oscillations, and thus dominate the dynamics.
For $p>1/2$, users migrate systematically from the less busy to the
busier server, until one is idle much of the time, and the other has
almost all clients in its queue. 

\begin{figure}
  \epsfxsize= 0.75 \columnwidth
  \epsffile{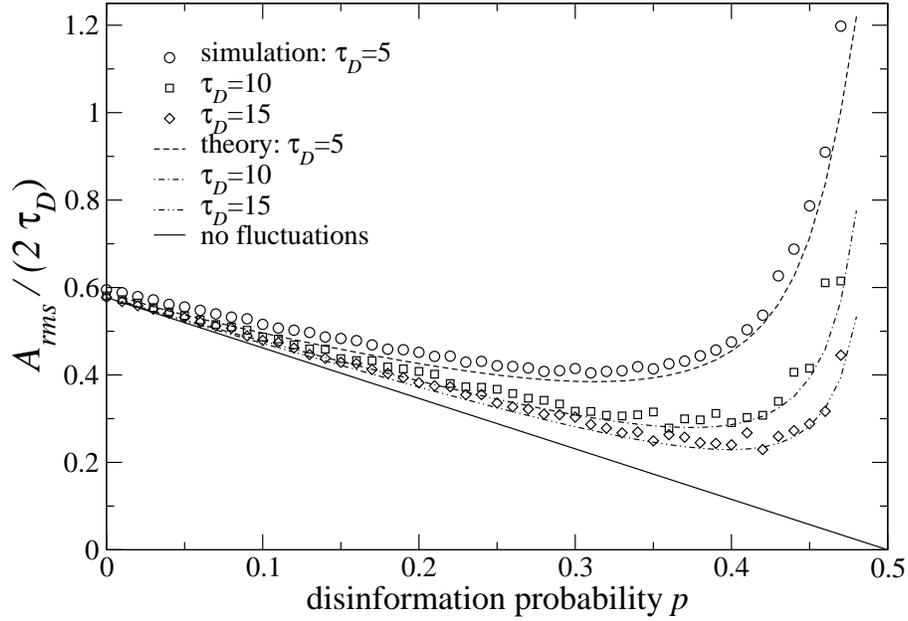}
  \epsfxsize= 0.75 \columnwidth
  \epsffile{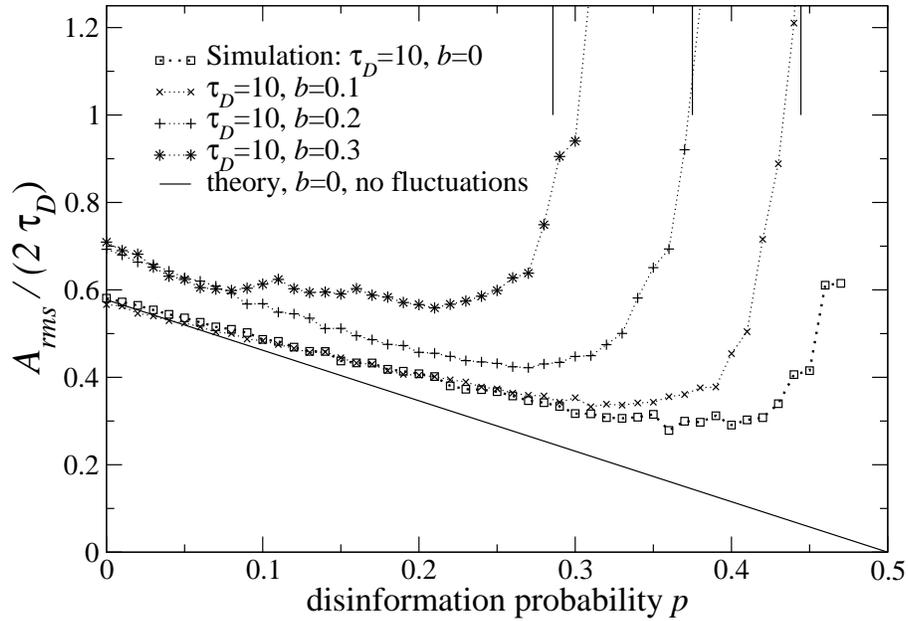}
  \caption{{\bf Top}: Global disinformation -- 
    the $y$-axis shows the root mean square
    difference in queue lengths, rescaled by the amplitude of the
    undisturbed system. 
    The theory for negligible fluctuations is given by
    Eq. (\ref{DEL-steady2}); agreement with simulations becomes better as
    absolute amplitudes increase. 
    The impact of fluctuations can be modeled using a discrete random walk. 
    \newline
    {\bf Bottom}: Adaptability -- if a percentage $b$ of clients always
    chooses the same server, disinformation disrupts coordination for
    values $p \lesssim (1-2b)/(2-2b)$, indicated by vertical lines for
    several values of $b$.}
  \label{DEL-disinfofig}
\end{figure}

The trade-off between reduced oscillations and increased fluctuations can
be seen in Fig. \ref{DEL-disinfofig}. Rather than measuring the amplitude of
oscillations, the root mean square $A_{rms} =
\langle A^2\rangle^{1/2}$ of $A(t)$ is
shown. For a pure triangle function of amplitude 
$a$, one gets $A_{rms}=a/\sqrt{3}$. 
For small $p$, the amplitude is reduced
linearly; for larger $p$,
fluctuations increase, dominating the dampened oscillations. 
When the amplitude of the undisturbed system is small, 
fluctuations have a large
impact. As the amplitude of oscillations gets larger, the 
impact of fluctuations becomes
smaller, and the value of $p$ where fluctuations dominate moves
closer to $1/2$.

Under the influence of randomness, $A$ performs a biased random walk: 
let us assume
that server $R_2$ is currently preferred. In each unit of time, $A/4$
increases by $1$ with probability $p^2$ (the two clients processed 
both go to $R_1$),
stays constant with probability $p(1-p)$, and decreases by 1 with probability 
$(1-p)^2$ (both clients move from $R_1$ to $R_2$). To reproduce quantitatively
the effects of fluctuations, one can numerically average $A^2$ over such a 
random walk
that takes place in two phases: the first phase lasts until $A=0$ is
reached; the second takes another $\tau_D$ steps until the direction of the
bias is reversed. The probability distribution of $A$ at the beginning of
the half-period has to be chosen self-consistently
such that it is a mirror-image of the probability distribution at the end;
the proper choice for $A/4$ is a Gaussian restricted to positive values 
with mean $(1-2p)\tau_D$ and variance $2 p(1-p)\tau_D$.
The numerical results are shown in Fig. \ref{DEL-disinfofig};
they agree well with values from the simulation. Note that in the above
treatment, we neglected complications like multiple crossings of the $A=0$
line. 

{\em Adaptability -- } Another aspect that determines what degree of
disinformation should be
chosen is adaptability of the system. 
The reason to have public information about the servers in the
first place is to be able to compensate for changes in the environment:
e.g., one server might change its data throughput due to hardware problems, or
some clients might, for whatever reason, develop a strong preference for
one of the servers. If the other agents can respond to these changes,
global coordination remains good; if their ability to understand 
the situation is too distorted by disinformation, global efficiency
suffers. 

Let us assume that
a fraction $b$ of agents decide to send their requests to server 1,
regardless of the length of queues. Under global disinformation,
out of the fraction $1-b$ that is left,
another fraction $d$ will send their requests to the wrong server, 
and yet another fraction $d$ is needed to compensate
for that. So a fraction $(1-b)(1-2d)$ is left to compensate for the action
of the biased group, and the maximum level of disinformation that still
leads to reasonable coordination is $p=(1-2b)/(2-2b)$ -- larger levels lead
to large differences in queue length, and finally to loss of data throughput by
emptying a queue.
That estimate is
confirmed by simulations (see Fig. \ref{DEL-disinfofig}). Similar arguments
hold if the preferences vary slowly compared to oscillation times.
On the other hand, if the preferences
of the biased agents oscillate in time with a period 
smaller than the delay time, they average out and have little effect on the
dynamics.

A similar argument also applies if the servers have different capacity. Let us
say $R_1$ has a processing time $\tau_{P1}$, whereas $R_2$ has
$\tau_{P2}> \tau_{P1}$. A fraction $f=\tau_{P2}/(\tau_{P1}+\tau_{P2})>1/2$ 
of clients should choose $R_1$ -- if $p$ is smaller than $1-f$, the queue of
$R_1$ will not become empty; otherwise it will.

{\em Individual rejection --} 
Even if servers cannot influence the public information on queue status,
they can influence the behavior of clients directly: they claim they are
not capable of processing a request, and reject it -- let us say, with a
constant probability $r$. Compared to global disinformation, 
a new possibility  arises that a request bounces back and forth several times
between the servers, but that adds nothing new in principle: the fraction  
of requests that
end up at the server that they tried at first is $(1-r) + r^2(1-r) +
r^4(1-r)+\dots = 1/(r+1)$, whereas a fraction $r/(r+1)$ will be processed
at the other server. This is equivalent to setting $p=r/(1+r)$ in the
``global disinformation'' scheme, and gives equivalent results.
Choosing $r$ close enough to $1$ reduces the amplitude of oscillations 
dramatically; however, each message is rejected a large number of times 
on average, generating large amounts of extra traffic. 

{\em Load-dependent rejection --} Rather than setting a constant rejection
rate, it seems intuitive to use a scheme
of load-dependent rejection (LDR), in which $r_i$ depends on the current length
of the queue. This is being considered for preventing the impact of single
server overload in Ref. \cite{Braden:Recommendations}.
 For example, let us consider
the case where $r_i =  c N_i$ if  $c N_i<1$,  and $1$ otherwise, with some
appropriately chosen constant $c$.
The analysis from the ``indiviual rejection'' section 
can be repeated with the additional
slight complication of two different rejection rates $r_1$ and $r_2$. 
A fraction $(1-r_1)/(1-r_1 r_2)$ of agents who initially try server 1 ends
up being accepted by it, whereas a fraction $r_1(1-r_2)/(1-r_1 r_2)$
finally winds up at server 2, and vice versa for clients who attempted $2$
first. Combining the resulting delay-differential equations for 
$N_1$ and $N_2$ into one for $A$, one obtains
\be
\frac{dA}{dt} = \frac{2}{1-r_1r_2}( \Theta(-A(t-\t_D))(1-2r_1+ r_1r_2) -
\Theta(A(t-\t_D))(1-2r_2+r_1r_2)). \label{DEL-steady4}
\ee
We can now substitute the load-dependent rates. 
We write them as follows: $r_1 = \bar{r} + c' A$, $r_2 = \bar{r} - c' A$, with 
$\bar{r}= (r_1+r_2)/2$ and $c' = c/2$.

\begin{figure}
  \epsfxsize= 0.9 \columnwidth
  \epsffile{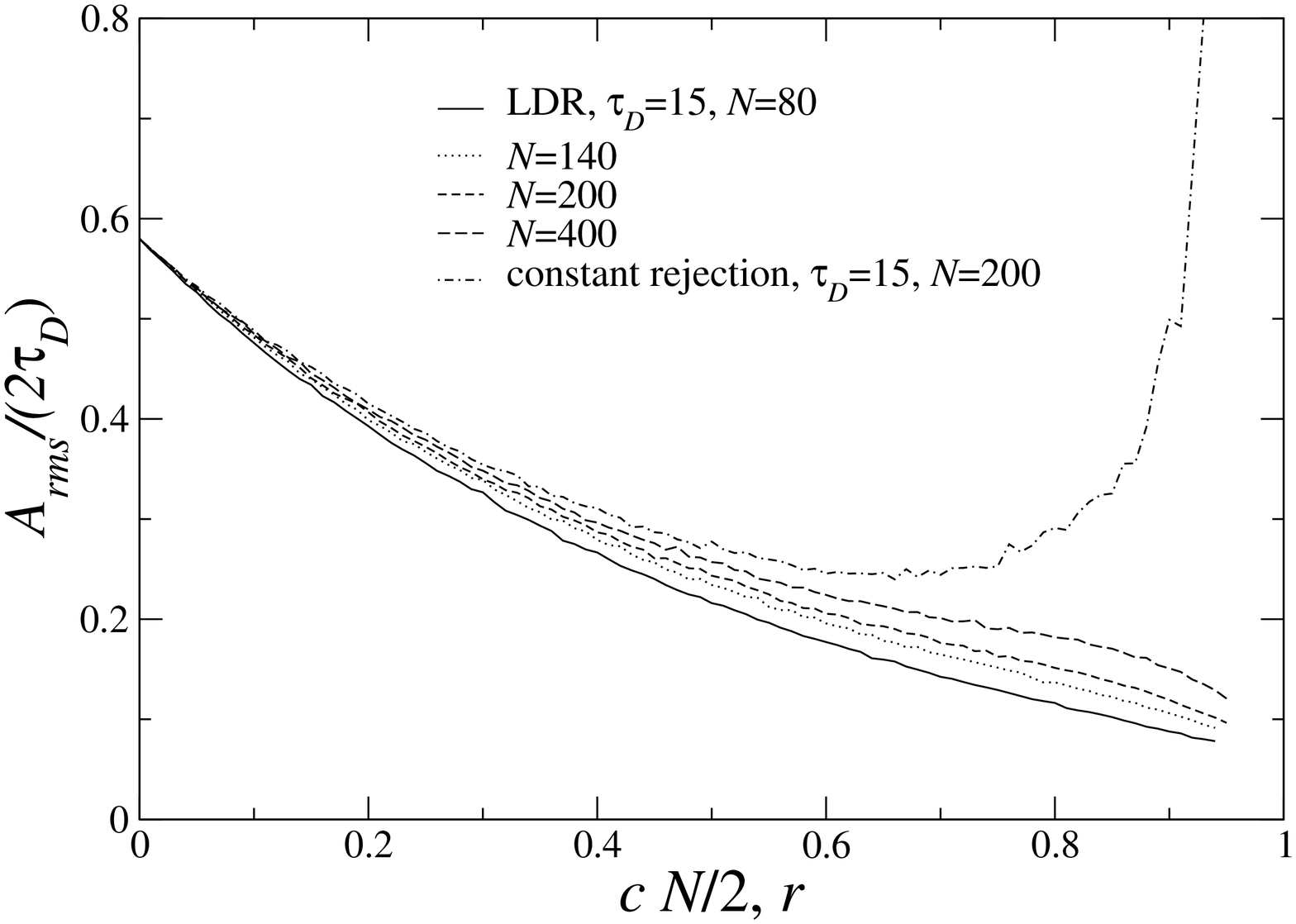}
  \caption{Load-dependent rejection: compared to rejection with a constant
    rate, LDR is more efficient at suppressing both oscillations and
    fluctuations. The former becomes
    more pronounced if $2\tau_D$ is not small compared to $N$.}
  \label{DEL-loaddep}
\end{figure}

For small amplitudes $A$ relative to the total number of
players $N$, the deviation from $\bar{r}$ does
not play a significant role, and it is $\bar{r}$ that determines behavior,
yielding the same results as a constant rejection rate.
For larger relative amplitudes, the oscillations are no longer pure triangle
waves, but have a more curved tip. Figure \ref{DEL-loaddep} shows $A_{rms}$
for load-dependent rejection, compared to constant-rate rejection with
$r=\bar{r}$. These nonlinear effects make LDR more efficient at suppressing
oscillations, at least if $2 \tau_D$ is not small compared to $N$. They
also provide for a restoring force that suppresses fluctuations effectively.
It follows that LDR is better at
improving data throughput in parameter regimes where servers empty out,
which is confirmed by simulations.  

We note one problem with LDR: if $c>2/N$, both servers have the maximal
number $1/c < N/2$
clients in their queue most of the time, 
while the rest of clients are rejected with probability 1 from both servers. 
For effective operation, this means that the constant $c$ in LDR should chosen
smaller than $2/N$, which requires knowledge of $N$.

{\em Discussion} -- We have introduced a model for the coordination of
clients' requests for service from two equivalent servers, found the
dependencies of the resulting oscillations on the parameters of the model,
and determined when and how these oscillations decrease data throughput.

We have then suggested a number of server-side ways to dampen the
oscillations, which involve purposely spreading wrong information about
the state of the servers. All of these schemes can achieve an improvement,
showing that the presence of faulty or incomplete information can be
beneficial for coordination.
The margins for improvement are higher in the regime of large numbers, when
the amplitude is on
the order of many tens or hundreds rather than a few individuals -- in the
latter case, increased fluctuations can outweigh the benefits of reduced 
oscillation.
While some disinformation generally improves performance, monitoring of the
average load and amplitude is necessary to choose the optimal degree of
disinformation. 

The basic ingredients of the server-client scenario (delayed public
information and minority-game-like structure) appear in many
circumstances. One can think of traffic guidance
systems that recommend one of two alternative routes, stock buying 
recommendations in magazines, career recommendations by employment centers,
and others. Exploring the impact of disinformation on
these problems is certainly worthwhile.


\begin{thebibliography}{10}
\expandafter\ifx\csname natexlab\endcsname\relax\def\natexlab#1{#1}\fi
\expandafter\ifx\csname bibnamefont\endcsname\relax
  \def\bibnamefont#1{#1}\fi
\expandafter\ifx\csname bibfnamefont\endcsname\relax
  \def\bibfnamefont#1{#1}\fi
\expandafter\ifx\csname citenamefont\endcsname\relax
  \def\citenamefont#1{#1}\fi
\expandafter\ifx\csname url\endcsname\relax
  \def\url#1{\texttt{#1}}\fi
\expandafter\ifx\csname urlprefix\endcsname\relax\def\urlprefix{URL }\fi
\providecommand{\bibinfo}[2]{#2}
\providecommand{\eprint}[2][]{\url{#2}}

\bibitem[{\citenamefont{Smith}(1904)}]{Smith:Inquiry}
\bibinfo{author}{\bibfnamefont{A.}~\bibnamefont{Smith}},
  \emph{\bibinfo{title}{An Inquiry into the Nature and Causes of the Wealth of
  Nations}} (\bibinfo{publisher}{Methuen and Co., Ltd},
  \bibinfo{address}{London}, \bibinfo{year}{1904}).

\bibitem[{\citenamefont{Challet and Zhang}(1997)}]{Challet:Emerg.}
\bibinfo{author}{\bibfnamefont{D.}~\bibnamefont{Challet}} \bibnamefont{and}
  \bibinfo{author}{\bibfnamefont{Y.-C.} \bibnamefont{Zhang}},
  \bibinfo{journal}{Physica A} \textbf{\bibinfo{volume}{246}},
  \bibinfo{pages}{407} (\bibinfo{year}{1997}).

\bibitem[{\citenamefont{Challet and Marsili}(1999)}]{Challet:Phase}
\bibinfo{author}{\bibfnamefont{D.}~\bibnamefont{Challet}} \bibnamefont{and}
  \bibinfo{author}{\bibfnamefont{M.}~\bibnamefont{Marsili}},
  \bibinfo{journal}{Phys. Rev. E} \textbf{\bibinfo{volume}{60}},
  \bibinfo{pages}{R6271} (\bibinfo{year}{1999}).

\bibitem[{MGH()}]{MGHomepage}
\bibinfo{title}{Minority game homepage (with extensive bibliography)},
  \bibinfo{note}{{\tt http://www.unifr.ch/econophysics/minority}}.

\bibitem[{\citenamefont{Nakar and Hod}(2002)}]{Nakar:Semianalytical}
\bibinfo{author}{\bibfnamefont{E.}~\bibnamefont{Nakar}} \bibnamefont{and}
  \bibinfo{author}{\bibfnamefont{S.}~\bibnamefont{Hod}} (\bibinfo{year}{2002}),
  \bibinfo{note}{cond-mat/0206056}.

\bibitem[{\citenamefont{Reents et~al.}(2001)\citenamefont{Reents, Metzler, and
  Kinzel}}]{Reents:Stochastic}
\bibinfo{author}{\bibfnamefont{G.}~\bibnamefont{Reents}},
  \bibinfo{author}{\bibfnamefont{R.}~\bibnamefont{Metzler}}, \bibnamefont{and}
  \bibinfo{author}{\bibfnamefont{W.}~\bibnamefont{Kinzel}},
  \bibinfo{journal}{Physica A} \textbf{\bibinfo{volume}{299}},
  \bibinfo{pages}{253} (\bibinfo{year}{2001}).

\bibitem[{\citenamefont{Marsili and Challet}(2000)}]{Marsili:Trading}
\bibinfo{author}{\bibfnamefont{M.}~\bibnamefont{Marsili}} \bibnamefont{and}
  \bibinfo{author}{\bibfnamefont{D.}~\bibnamefont{Challet}}
  (\bibinfo{year}{2000}), \bibinfo{note}{cond-mat/0004376}.

\bibitem[{\citenamefont{Savit et~al.}(1999)\citenamefont{Savit, Manuca, and
  Riolo}}]{Savit:Comp.}
\bibinfo{author}{\bibfnamefont{R.}~\bibnamefont{Savit}},
  \bibinfo{author}{\bibfnamefont{R.}~\bibnamefont{Manuca}}, \bibnamefont{and}
  \bibinfo{author}{\bibfnamefont{R.}~\bibnamefont{Riolo}},
  \bibinfo{journal}{Phys. Rev. Lett.} \textbf{\bibinfo{volume}{82}},
  \bibinfo{pages}{2203} (\bibinfo{year}{1999}).

\bibitem[{\citenamefont{Klein and Bar-Yam}(2003)}]{Klein:Handling}
\bibinfo{author}{\bibfnamefont{M.}~\bibnamefont{Klein}} \bibnamefont{and}
  \bibinfo{author}{\bibfnamefont{Y.}~\bibnamefont{Bar-Yam}}, Handling Resource Use Oscillation in Open Multi-Agent Systems.
   \bibinfo{note}{aAMAS Workshop} (\bibinfo{year}{2003}).

\bibitem[{\citenamefont{Braden et~al.}(1998)\citenamefont{Braden, Clark,
  Crowcroft, Davie, Deering, Estrin, Floyd, Jacobson, Minshall, Partridge
  et~al.}}]{Braden:Recommendations}
\bibinfo{author}{\bibfnamefont{B.}~\bibnamefont{Braden}},
  \bibinfo{author}{\bibfnamefont{D.}~\bibnamefont{Clark}},
  \bibnamefont{et~al.}, Recommendations on Queue Management and Congestion Avoidance in the Internet.  \bibinfo{note}{{N}etwork
  {W}orking {G}roup} (\bibinfo{year}{1998}).

\end{thebibliography}
\end{document}